\documentclass{aa}  
\usepackage[switch]{lineno}
\linenumbers
\nolinenumbers

\usepackage{float}
\usepackage{graphicx}
\usepackage[version=4]{mhchem}
\usepackage[table,xcdraw]{xcolor}
\usepackage{colortbl}
\usepackage{txfonts}
\begin{document}

   \title{Self-consistent 1D Modelling of Jupiter's Upper Atmosphere as an Exoplanet Analogue}

   \subtitle{}

   \author{Nils-Martin Robeling
          \inst{1}
          \and
          Sudeshna Boro Saikia\inst{1}
          \and
          Gwena\"elle Van Looveren\inst{1}
          \and
          Ivan Stankovi\'c
          \inst{1}
          \and
          Simon Schleich\inst{1}
          \and
          Colin P. Johnstone\inst{2}
          \and
          Manuel G\"udel\inst{1} 
          \and
          Kristina Kislyakova\inst{1}
            }

   \institute{Department of Astrophysics, University of Vienna,
              T\"urkenschanzstrasse 17, A-1180 Vienna \email{nils-martin.robeling@univie.ac.at} \and International Institute for Applied Systems Analysis, Schlo\ss{}platz 1, A-2361, Laxenburg, Austria\\ 
              }

   \date{}

  \abstract{Jupiter's upper atmosphere provides a real-world laboratory for validating first-principles models of giant gaseous exoplanets and for constraining the key physical processes that govern them.}{We extended the 1D first-principles thermo-chemical planetary upper atmosphere model Kompot to simulate hydrogen-rich atmospheres of giant exoplanets and benchmarked it against the archetype giant planet Jupiter.}{We modelled a longitudinal and latitudinal average thermal and chemical profile of Jupiter's upper atmosphere by solving the equations of energy balance, photochemical kinetics, hydrodynamics, and vertical transport in a 1D radial atmospheric grid. The thermal properties of the upper atmosphere were determined by a balance between heating from solar X-ray and ultraviolet (XUV), and infrared (IR) radiation, thermal conduction, Joule heating, and radiative cooling due to H$_3^+$ and {CH$_4$}. The model results were compared with Jupiter observations from Galileo, JUNO, and other instruments. The simulation results were also used as input in the radiative transfer module of the TauREx code to simulate Jupiter's infrared transmission spectrum.}{Our model reproduces Jupiter's observed upper-atmospheric thermal structure and observed CH$_4$ volume mixing ratios. The chemical abundances of other gases also exhibit strong agreement with observations and existing photochemical models. Heating and cooling results show that Joule heating is the dominant heating source throughout most of the upper atmosphere. The transmission spectrum indicates the rich presence of CH$_4$ in the near- and mid-IR wavelength ranges.}{Our physically consistent model framework for Jupiter's upper atmosphere provides a validated baseline for future self-consistent simulations of the thermal and chemical structure of a diverse population of hydrogen-rich exoplanet atmospheres.}

    \titlerunning{Self-consistent 1D
Modelling of Jupiter’s Upper Atmosphere}
    \authorrunning{Nils-Martin Robeling et al.}
   \maketitle


\section{Introduction}
Understanding the processes shaping the thermal and chemical structure of Jupiter’s atmosphere is essential, not only for interpreting observations of our Solar System’s gas giants but also for modelling the atmospheres of close-in giant exoplanets, such as hot Jupiters. Although hot Jupiters and Jupiter have different temperatures and orbital distances, Jupiter is the closest analogue in mass, radius, and composition for which we have reliable measurements of temperature and chemical abundances. Recently, the \emph{James Webb} Space Telescope \citep[JWST;][]{jimothywebbs} has detected molecules such as CH$_4$ \citep{bell_methane_2023}, SO$_2$ \citep{tsai_photochemically_2023}, CO$_2$ \citep{jwst_transiting_exoplanet_community_early_release_science_team_identification_2023}, and CO \citep{grant_detection_2023} in the atmospheres of exo-gas giants. 
   
Even though these detections provide insight into exoplanet atmospheric chemistry, observations alone cannot provide significant insight into the thermal profile and energy budget \citep{schleich_knobs_2024}. Photochemical kinetics models \citep[e.g.][]{moses_photochemistry_2005, tsai_vulcan_2017, knizek_full-atmosphere_2025} are widely used to model the chemical profile of such planets, where the thermal profile is generally an input parameter. Radiative-convective equilibrium (RCE) models \citep[e.g.][]{amundsen_accuracy_2014, baudino_interpreting_2015, tremblin_fingering_2015} are used to describe the steady-state temperature structure where radiative cooling and convective heating balance in an atmospheric column. RCE models provide a computationally efficient framework for the lower-atmospheric temperature structure, but they are blind to effects such as heating from exothermic chemical reactions, disequilibrium chemistry, radiative cooling, and thermal conduction. Hence, models that solve the coupled hydrodynamics, chemical kinetics, and energy balance equations, benchmarked against Jupiter observations, can be very effective. 

Jupiter’s upper atmosphere has, over the past decades, been observationally explored by spacecraft and Earth-based telescopes, but the details of its energetics and composition still raise many unresolved questions, especially since these quantities also vary spatiotemporally \citep{romani_temporally_2008,kim_temporal_2020,munoz_upper_nodate}. More specifically, the H$_3^+$ temperature of Jupiter can vary spatially by ~600 K \citep{roberts_spatiotemporal_2025}, and the number densities of hydrocarbons such as C$_2$H$_2$ and C$_2$H$_6$ deviate by up to an order of magnitude, depending on the longitude and latitude. In the upper atmosphere, above the stratospheric haze layer, we face uncertainties and significant variations due to the banded cloud structures and high-speed jet streams and waves that dominate the atmospheric dynamics of Jupiter \citep{kaspi_jupiters_2018}. In addition to the vertical structure, the effects from deep atmospheric circulation and waves are poorly mapped and difficult to quantify \citep{garcia-melendo_dynamics_2011}. 

Over the past few decades, a wide array of models has been developed to decipher the structure and energetics of Jupiter’s upper atmosphere. Early 1D photochemical and ionospheric models \citep[e.g.][]{Strobel1973, gladstone_hydrocarbon_1996, perry_chemistry_1999} established the basic roles of hydrocarbon photolysis and ion–neutral chemistry but generally relied on prescribed temperature profiles and simplified diffusion schemes. Other efforts have increasingly sophisticated chemistry and vertical mixing \citep[e.g.] []{moses_photochemistry_2005}. The thermal structure of Jupiter's upper atmosphere has long been a perplexing problem, as it was too hot for models to be explained solely by solar heating. Several theories, such as heating from gravity waves \citep{young_gravity_1997, matcheva_heating_1999}, dissipating and depositing heat, or particle precipitation and auroral heating, are widely discussed \citep{horanyi_precipitation_1988}. However, it is stated that, for gravity wave heating, the observed quantities are insufficient to achieve the required heating of the order of magnitude.  At the same time, auroral, magnetic, and ionospheric models have highlighted the importance of Joule heating and particle precipitation in shaping the thermal structure at high latitudes \citep{ nishida_joule_1981,grodent_selfconsistent_2001}.  Despite this progress, models attempting to compute both the chemical and the thermal profiles self-consistently are rare. \citet{grodent_selfconsistent_2001} present a self-consistent model of the upper atmosphere, from 1 mbar to the exobase, in the auroral regions. 
However, an average self-consistent radial model of Jupiter's thermal and chemical profile as an exoplanet analogue does not yet exist. This gap becomes increasingly relevant as analogous processes are used to explain the energetics of hot Jupiter atmospheres, yet benchmarked models for Jupiter-like conditions remain limited.

While all the above-mentioned models only capture one dimension, general circulation models (GCMs) can serve as powerful tools for capturing the 3D connections between radiative transfer and large-scale dynamics. The primary advantage of GCMs lies in their ability to simulate horizontal transport. Specifically, they can capture the intense zonal jets and meridional heat redistribution that dominate the upper atmospheric temperature profile of Jupiter’s non-auroral regions. For Jupiter, there are multiple GCMs that have effectively simulated the upper atmosphere and its temperature variations, such as \citet{majeed_processes_2005} or \citet{muller-wodarg_temperatures_2025}. However, incorporating highly detailed calculations of disequilibrium processes, such as comprehensive photochemistry networks and detailed cooling functions, remains computationally prohibitive in a 3D framework. Another caveat is that the regime of exoplanet GCMs usually does not stretch above the homopause, and detailed calculations of disequilibrium processes, such as photochemistry, are primarily done in 1D. While GCMs offer an inherently self-consistent treatment of coupled non-linear dynamics and major energetic inputs (e.g. solar and Joule heating), they trade off microphysical and chemical complexity to remain computationally feasible. This is where 1D models featuring highly detailed, self-consistent vertical physics and complex chemical networks can complement GCMs.

For this work, we extended the 1D atmospheric model Kompot, originally developed for rocky bodies in the Solar System, to gas-giant atmospheres. We set the simulation boundary above the cloud deck, and self-consistently simulated the average thermal and chemical structure of Jupiter's upper atmosphere over the pressure range $10^{-3}$ to the exobase, treating Jupiter as an exoplanet around a Sun-like star. Our model represents a globally averaged radial atmospheric structure rather than localised or time-dependent conditions. This paper is structured as follows. Section 2 describes the Kompot model, the other methods used, and the assumptions in this work. Section 3 discusses the results, and Section 4 provides the conclusions.

\section{Model}
The Kompot code is a 1D first-principles model that computes the thermal and chemical structure, as well as the Jeans escape rate, of planetary upper atmospheres. The model does this by solving the hydrodynamical equations coupled with the energy balance and chemistry. In this work, we define the upper atmosphere as the atmospheric layer above the stratospheric haze layer and cloud decks on Jupiter at $10^{-3}$ bar. As input, Kompot requires a boundary condition comprising molecular abundances, base temperature, eddy diffusion profile, and parameters relevant to Joule and XUV/IR heating, as further discussed in Sect. \ref{sec:input}. The model's output is a thermal and chemical profile that includes contributions from different heating and cooling sources. The Kompot code was originally developed for rocky planets in the Solar System. This work extends the model to gas giants. For more detailed information about the model, see \citet{johnstone_upper_2018,johnstone_extreme_2019, johnstone_young_2021}. Below, we discuss the key processes involved in Kompot and the modifications made to make it suitable for gas giant atmospheric modelling. This is followed by a discussion on the radiative transfer model used to generate a transmission spectrum from our Kompot simulation.
\begin{table*}[h!]
\caption{Model input conditions.}
\label{tab:parameters}
\begin{tabular}{cccc}
\hline

Parameter              & Value                                 & Reference                                             & Note                       \\ \hline
Eddy diffusion profile & See Ref. and Fig. \ref{fig:diff} [cm/s$^2$]                 & Model A and C from \citealt{moses_photochemistry_2005}                &               Includes $K_{zz} = 10^6 / 10^7$      \\ 
X(H$_2$)               & 8.35 $\times 10^{-1}$                 & \cite{moses_photochemistry_2005}                     & Lower boundary mixing ratio      \\ 
X(He)                  & 1.62 $\times 10^{-1}$                 & \cite{moses_photochemistry_2005}                     & Lower boundary mixing ratio      \\ 
X(CO)                  & 4.42 $ \times 10^{-8}$                 & \cite{moses_photochemistry_2005}                     & Lower boundary mixing ratio      \\ 
X(N$_2$)               & 3 $\times  10^{-5}$                    & \cite{moses_photochemistry_2005}                     & Lower boundary mixing ratio      \\ 
X(CH$_4$)              & 2.1 $\times 10^{-3}$                  & \cite{moses_photochemistry_2005}                    & Lower boundary mixing ratio      \\ 
X(C$_2$H$_2$)              & 2 $\times 10^{-7}$                  & \cite{moses_photochemistry_2005}                    & Lower boundary mixing ratio     \\ 
X(C$_2$H$_4$)              & 4 $\times 10^{-9}$                  & \cite{moses_photochemistry_2005}                    & Lower boundary mixing ratio      \\ 
X(C$_2$H$_6$)              & 8 $\times 10^{-6}$                  & \cite{moses_photochemistry_2005}                    & Lower boundary mixing ratio      \\ 
n                      & 4.36$\times  10^{16} $ {[}cm$^{-3}${]} & Galileo probe measurement \citep{yelle_structure_2001}    & Lower boundary number density \\ 
T                      & 165 {[}K{]}                           & Galileo probe measurement \citep{yelle_structure_2001}    & Lower boundary temperature    \\ 
$\Gamma_j$             & ~$\sim 10^{21}$ {[}erg/s{]}    & \cite{muller-wodarg_temperatures_2025} & Total Joule heating        \\ 
B                      & 4.17 {[}G{]}                          &  \cite{connerney_new_2018}               & Magnetic field strength    \\ 
$\theta_{Z}$                      & 15-75 (60) {[}$\circ${]}                           & -     & Zenith angle  (main models)  \\ \hline
\end{tabular}
\end{table*}
\begin{figure}[h!]
    \centering
    \includegraphics[width=1.1\linewidth]{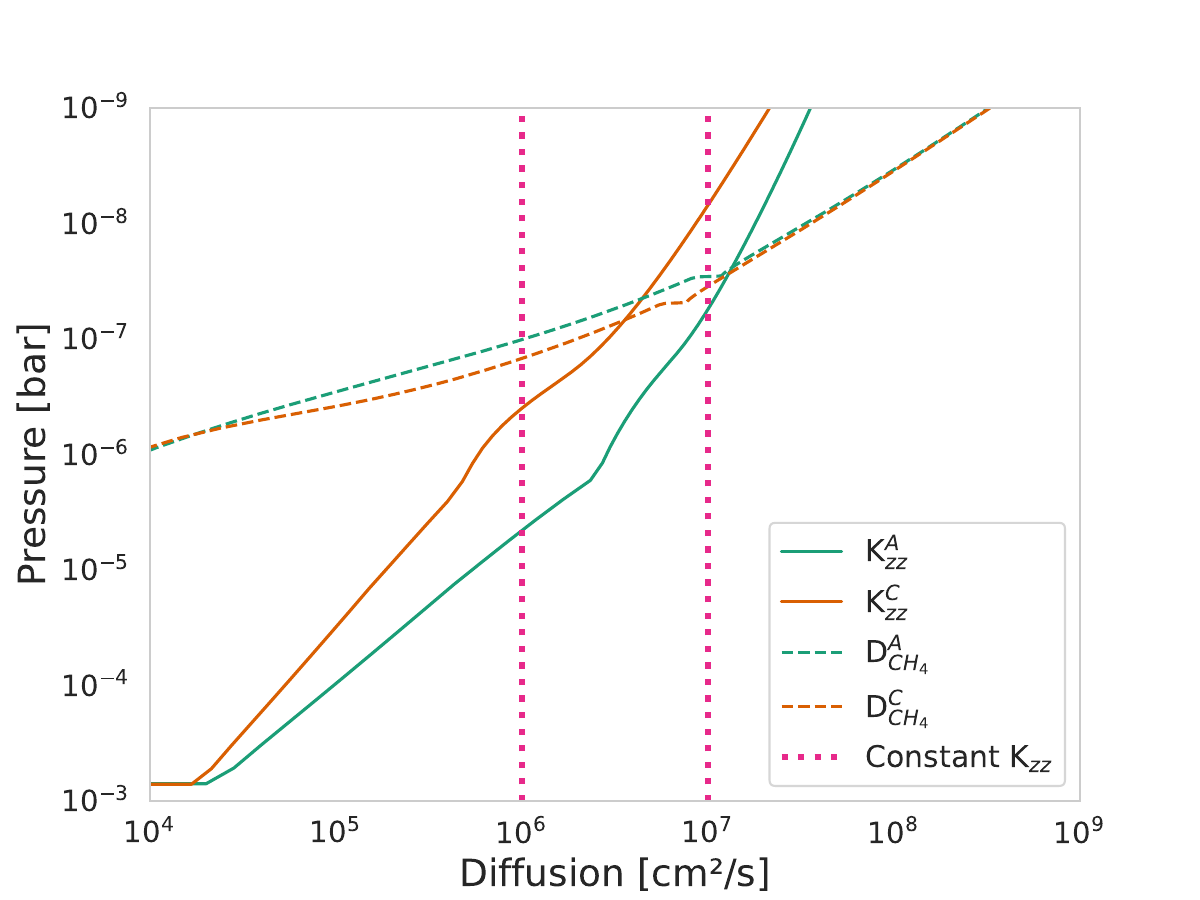}
    \caption{Molecular and eddy diffusion considered for the different models in this work. The shaded area represents the possibility space of Jupiter's homopause location with results from \cite{sinclair_spatial_2020} and \cite{rodriguez-ovalle_temperature_2024}.}
    \label{fig:diff}
\end{figure}
\subsection{Hydrodynamics}
While the Kompot code can simulate hydrodynamic atmospheres, in this work we considered the hydrostatic case. This means that in

\begin{equation}\label{hydroeq1}
    \frac{1}{\rho} \frac{\mathrm{~d} \rho}{\mathrm{~d} r}=-\frac{1}{T} \frac{\mathrm{~d} T}{\mathrm{~d} r}-\frac{g}{v_{0}^{2}}+\frac{1}{\bar{m}} \frac{\mathrm{~d} \bar{m}}{\mathrm{~d} r}-\frac{v}{v_{0}^{2}} \frac{\mathrm{~d} v}{\mathrm{~d} r},
\end{equation}
where $\rho$ is density in g/cm$^3$, $T$ is temperature in K, $r$ is radius in cm, $g$ is the gravitational acceleration in cm/s$^2$, $\bar{m}$ is the mean molecular mass of the gas in Da, $v$ is the bulk vertical advection speed in m/s, and $v^2_0 = k_BT/\bar{m}$. We set $v=0$. This effectively gives us the differential of the density structure as
\begin{equation}\label{hydroeq2}
    \frac{1}{\rho} \frac{\mathrm{~d} \rho}{\mathrm{~d} r}=-\frac{1}{T} \frac{\mathrm{~d} T}{\mathrm{~d} r}-\frac{g}{v_{0}^{2}}+\frac{1}{\bar{m}} \frac{\mathrm{~d} \bar{m}}{\mathrm{~d} r},
\end{equation}
which is then integrated from the lower boundary, where the density is calculated based on the input number density, up to the exobase. The exobase is dynamically calculated and is defined as the boundary before a transition into the collisionless regime, where the mean-free path of a particle becomes equal to the scale height of the atmosphere
\begin{equation}
    \frac{1}{\sqrt{2}\sigma n} = \frac{k_BT}{\bar{m}g},
\end{equation}
where $\sigma$ is the collisional cross-section area of molecular hydrogen, $n$ is the number density, and $k_B$ is the Boltzmann constant. This is also known as the altitude at which the Knudsen number, the ratio of the mean-free path to the scale height, equals 1. Above the exobase, the particles no longer follow a Maxwellian velocity distribution, and Jeans escape becomes increasingly relevant.

\subsection{Chemistry}
The chemical network used in Kompot includes neutral, ionic, and photochemical reactions, as well as hydrocarbon chemistry up to C$_2$H$_6$. Reaction rate coefficients are primarily taken from the Kinetic Database for Astrochemistry (KIDA) \citep{wakelam_kinetic_2012}, supplemented by additional sources and other networks such as those of \cite{yung_evidence_2009}, \cite{garcia_munoz_physical_2007},  \cite{richards_reexamination_2011}, \cite{fox_chemistry_2015}, and VULCAN \citep{tsai_vulcan_2017} where necessary. We use cross-sections for XUV photochemistry from PHIDRATES \citep{herbst_update_1985,huebner_photoionization_2015}.

The coupled set of continuity equations governing the transport and transformation of chemical species is solved using a Rosenbrock-type semi-implicit integrator \citep{sandu_benchmarking_1997}. This type of solver is well-suited to stiff systems, characterised by the coexistence of fast radical reactions and slow photolytic or recombination processes. 
\begin{figure}
    \centering
    \includegraphics[width=1.1\linewidth]{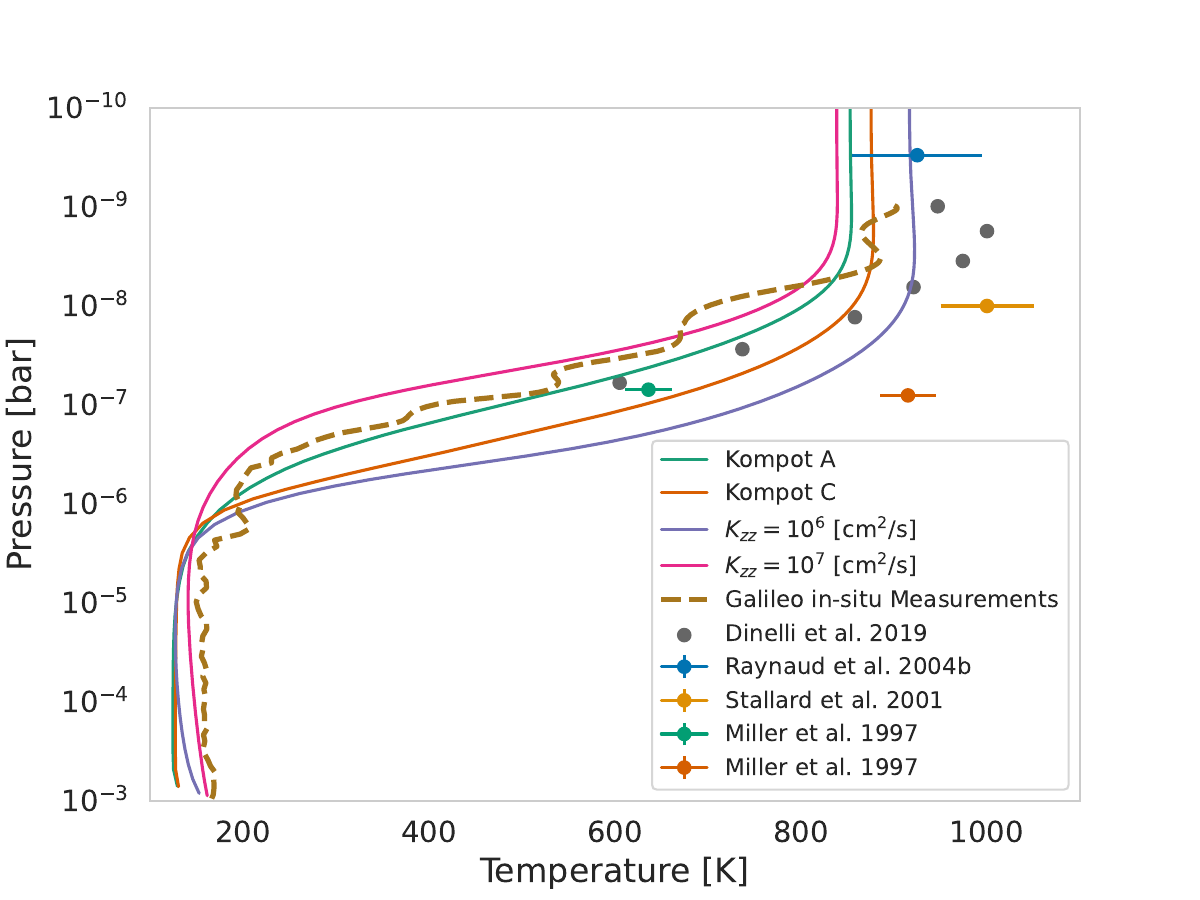}
    \caption{Thermal profiles of our four models from Kompot. Model A and C use the corresponding K$_{zz}$ from \cite{moses_photochemistry_2005}, and the other two models use constant values. Here we compare them with in situ measurements and with different measurements from Earth-based telescopes, as well as with one in situ measurement. The in situ measurement is from the Galileo probe, entering Jupiter's atmosphere at a 5 $\mu$m hot spot. The markers with error bars come from different ground-based telescopes measuring Jupiter's auroral regions. Raynaud et al. used the CFHT-FTS (Canada-France-Hawaii Telescope-Fourier transform spectrometer) in the northern hemisphere at 150$^\circ$-170$^\circ$ System III longitude \citep{raynaud_spectro-imaging_2004}, Stallard et al. used the CSHELL spectrometer on the NASA-IRTF (Infrared Telescope Facility) \citep{stallard_dynamics_2002}, and Miller et al. used the UKIRT (United Kingdom Infrared Telescope) \citep{miller_mid--low_1997}. The grey scatter points from \cite{dinelli_junojirams_2019} are from the JIRAM (Jovian Infrared Auroral Mapper) on JUNO via retrievals.}
    \label{fig:mainPT}
\end{figure}
\begin{figure*}
    \centering
    \begin{tabular}{cc}
        \includegraphics[width=0.5\textwidth]{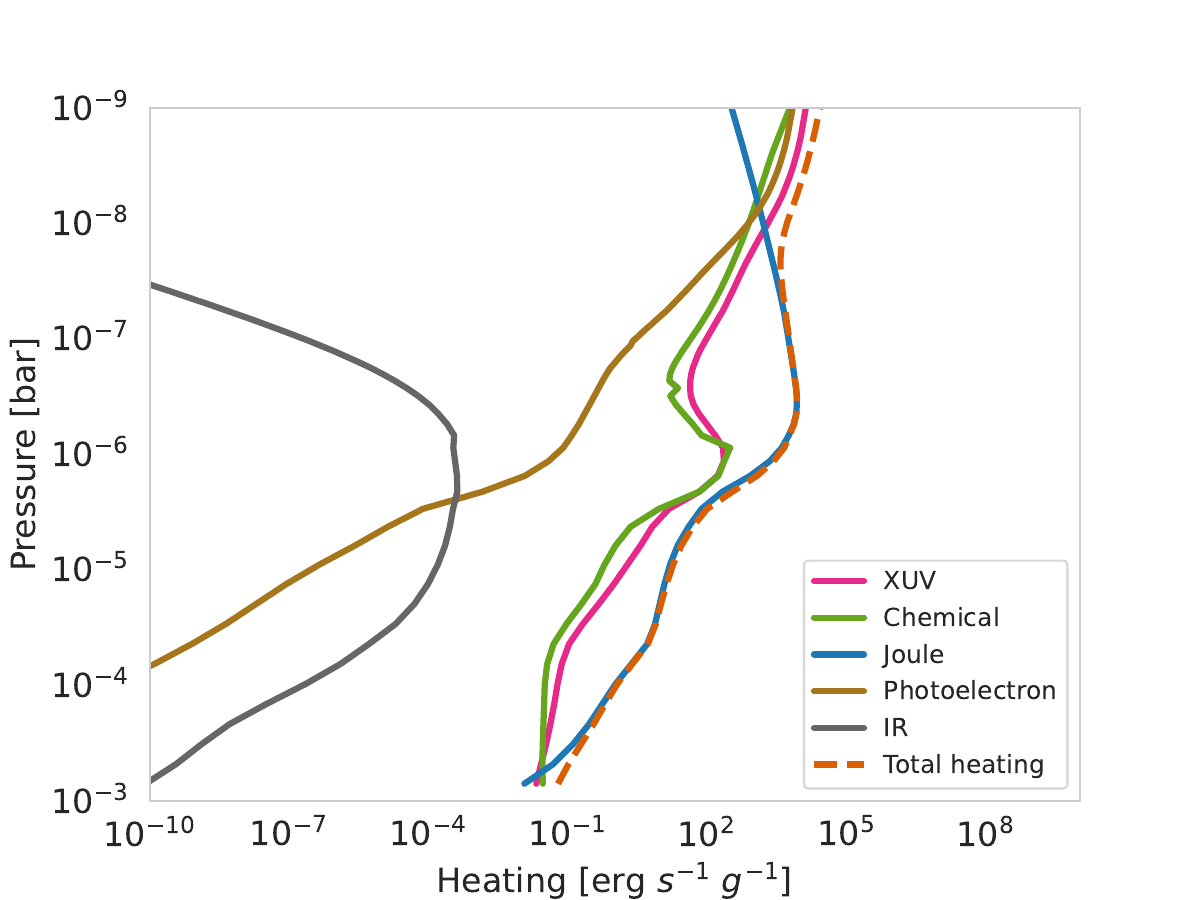} &
        \includegraphics[width=0.5\textwidth]{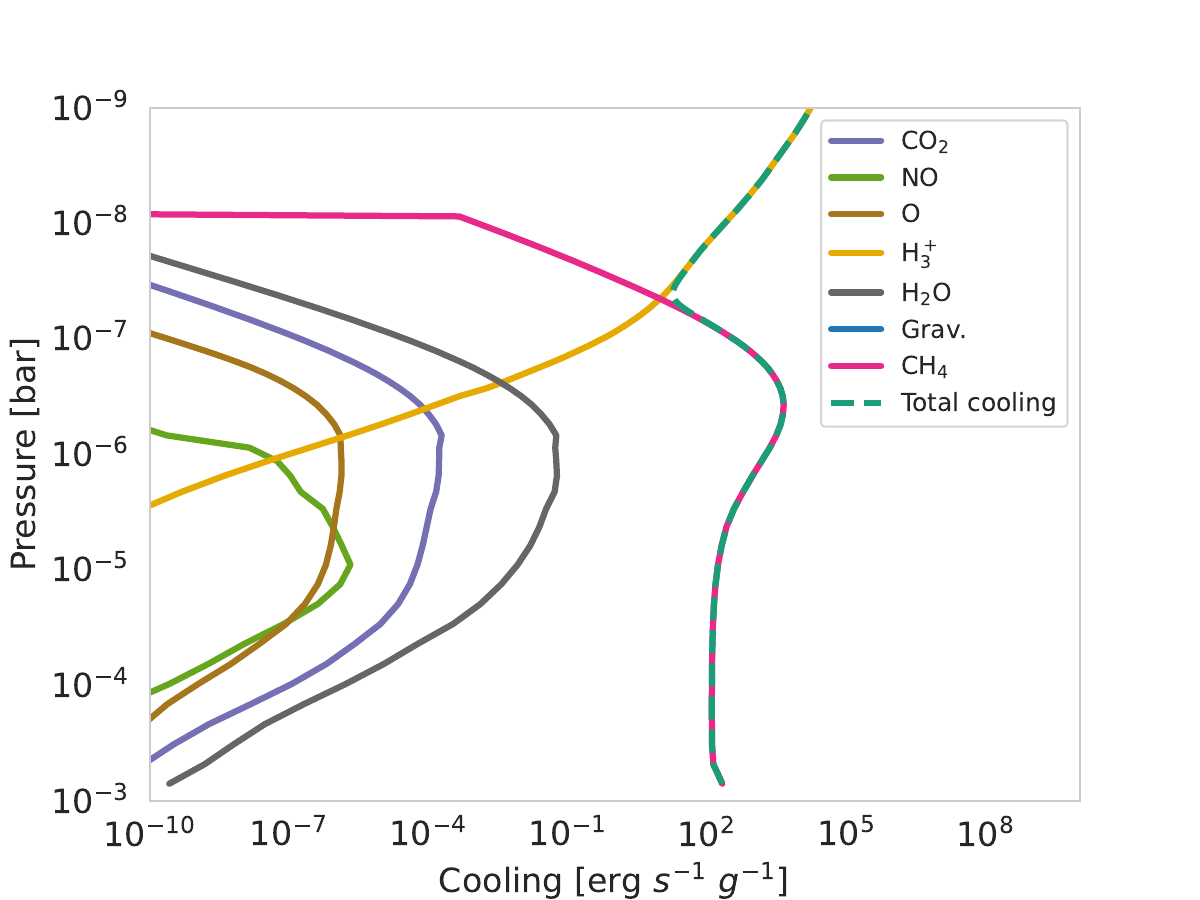}
    \end{tabular}
    \caption{Total heating and cooling for Model A shown together with the individual processes that contribute towards the total. Left: The constituents of heating together with total heating. Right: The dominant cooling terms are shown together with the total cooling. Please note that gravitational cooling is too low to be present in the plot.}
    \label{fig:HCcomps}
\end{figure*}
\subsection{Energy balance}\label{sec:energy}
    A steady-state model atmosphere requires the balance of heating, cooling, and thermal conduction. In this section, we describe the key processes affecting the thermal structure in our 1D model; for a more detailed description, refer to \citet{johnstone_upper_2018} or references herein. The only free parameters in the heating module are the magnetic field and total Joule heating input energy.
    \subsubsection{Heating}\label{sec:heating}

    The heating rate from XUV radiation is obtained using a simplified radiative transfer model. We assign the energy not used in photoreactions from the XUV radiation field to the gas as heat. For each atmospheric layer $i$, the heating contribution is calculated by summing over all absorption reactions and photon energy bins. The heating contribution accounts for the excess photon energy above the ionisation threshold, weighted by the local photon flux, the bin width in energy space, and the corresponding reaction cross sections, yielding the volumetric heating rate $Q_{XUV,i}$ of layer $i$. 

    Heating from photoelectrons, via photoionisation, is calculated following the approach of \cite{schunk_electron_1978}. The electron temperature characterises the thermal background, while the photoelectron velocity distribution is obtained from their kinetic energy. The first energy bin where the photoelectron flux exceeds the thermal Maxwell-Boltzmann flux is identified as the threshold for heating. This is where the photoelectron population starts dominating over the background thermal electrons. Above this threshold, the energy loss rate of non-thermal electrons is given by Eqs. 9 and 11 of \cite{schunk_electron_1978}. The corresponding heating rate per atmospheric layer $i$ is then computed using their Eq. 7: 
    
    \begin{align}
    \Gamma_{PE,i} &= \sum_j P_{e,i}(E_j) \frac{dE}{dx}(E_j)\, dE_{PE}(E_j) \notag \\
    &\quad + \left(E_{\min} - E_{\mathrm{therm}}\right)\frac{dE}{dx}(E_{\min}) P_{e,i}(E_{\min}),
    \end{align}
    where $P_e(E_x)$ is the photoelectron flux spectrum, $dE/dx$ is the energy loss rate, and $E_{\min}$ denotes the first energy bin above the thermal background.
    
    Joule heating in our model is a scaled free parameter following the formalism of \citet{Foster_joule1983}. While total Joule heating, $Q_j$, is a free parameter specified in the initial conditions, the fraction deposited in each layer must still be calculated. We do this by first finding the
    ion cyclotron frequencies as
    \begin{equation}\label{eq:omegac}
        \omega_{c_i} = \frac{e_iB}{m_ic},
    \end{equation}
    where $e_i$ is the electric charge of species $i$, $B$ is the magnetic field strength, $c$ is the speed of light, and $m_i$ is the mass of the species. We then move on to find the momentum transfer collisional frequencies for resonant charge exchange as
    \begin{equation}\label{eq:nuin}
        \nu_{in} = \frac{8}{3\sqrt{\pi}}n_n \Biggr[ \frac{2k(T_i + T_n)}{m_i}\Biggr]^{1/2}[A' + 3.96B' - B'\text{log}_{10}(T_i + T_n)]^2,
    \end{equation}
    where $n_n$ is the neutral density, $T_i$ is the ion temperature, $T_n$ is the neutral temperature, $m_i$ is the mass of the ion, and $A'$ and $B'$ are constants from \cite{schunk_ionospheres_nodate} that depend on the considered species.
    
    We also add the non-resonant collision frequencies for ion-neutral pairs with $\nu_{in} = C_{in}n_n$, where C is a numerical coefficient from Table 4.4 in \cite{schunk_ionospheres_nodate}. Together with the collision frequencies, we can find the conductivity of the ions as
    \begin{equation}\label{eq:sigmai}
        \sigma_i = \frac{n_ie_i^2}{m_i\nu_i}.
    \end{equation}
    We put Eqs. \ref{eq:omegac}, \ref{eq:nuin}, and \ref{eq:sigmai} into the following equation to find the Pedersen conductivity via the expression,
    \begin{equation}\label{eq:pedersen}
        \sigma_P = \sum \sigma_i \frac{\nu_i^2}{\nu_i^2 + \omega^2_{c_i}},
    \end{equation}
    where $i$ is the considered ion, $\sigma_i$ is the ion conductivity, $\nu_i$ the collision frequency in total for one ion. The above equation is a simplification since the electron contribution is removed because $\nu_e \ll \omega_{c_e}$ in most cases, since electrons are strongly tied to the magnetic field and the collisions are too infrequent to disrupt their gyration, thus keeping terms that are only of order $\nu_e/\omega_{c_e}$. After retrieving the Pedersen conductivity, we form the expression for Joule heating,
    \begin{equation}
        Q_j = \sigma_PE^2,
    \end{equation}
where $Q_j$ is the Joule heating per cell and E is the electric field strength. To estimate the spatial distribution of the Joule heating, we set E=1, following which we only require the shape of the Pedersen conductivity profile to obtain the distribution of the Joule heating across all cells. To obtain the actual energy values per bin we first sum up the contribution across all cells as
    \begin{equation}
        Q_{rad} = \int^{R_{pl}} 4\pi \sigma_P(r) r^2 dr,
    \end{equation}
    to find the final estimate of $Q_j$ per cell by scaling the free parameter, total heating input, to local heating input 
    \begin{equation}
        Q_j \approx\sigma_P\frac{Q_{j,input}}{Q_{rad}},
    \end{equation}
    where $Q_{j,input}$ is the total input Joule heating to the atmosphere.
    Effectively, this calculates a Pedersen conductivity profile throughout the atmosphere and then distributes the model input total Joule heating, which is a free parameter, over it. Since the Joule heating profile strongly follows the shape of the Pedersen conductivity profile, we have chosen to include only the Joule heating profile.
    
    Finally, the heating from chemical reactions is calculated by summing the net energy released or absorbed for exothermic and endothermic reactions, each multiplied by the reaction rate per layer, using formation enthalpy values.

    Although the Kompot code contains multiple sources of heating, it is imperative to mention that it does not include heating from the particle precipitation prevalent in the auroral regions of Jupiter \citep{rodriguez-ovalle_temperature_2024}, and is thus not suitable for simulating these hotter regions specifically, but rather the horizontally averaged profile.

\subsubsection{Cooling}\label{sec:cooling}
    The cooling of Jupiter's upper atmosphere is dominated by two major sources: line cooling from H$_3^+$, and radiative cooling from CH$_4$. These two cooling effects operate and dominate in distinct regions. Well below the homopause, where CH$_4$ does not experience heavy photolysis, radiative cooling from CH$_4$ dominates, since no other strong cooling sources are available. Line cooling from H$_3^+$ is first effective above the homopause, where H$_3^+$ becomes available through interactions with solar photons. The H$_3^+$ line cooling is handled according to \cite{miller_cooling_2013} using the coefficients available in their Table 5, which are then multiplied by a scale factor to account for non-LTE effects and the density of H$_3^+$. Radiative cooling from CH$_4$ uses ExoMol \citep{tennyson_2024_2024} data to fit a function between 1 and 1500 K to the cooling in line with \cite{tennyson_radiative_2016}. This does not account for non-LTE effects, which is not a big problem for the model since CH$_4$ cooling dominates below the homopause, where LTE is a good approximation; however, this approximation may make the model slightly cooler. For a more detailed description, refer to \cite{tennyson_radiative_2016} or Boro Saikia et al. (in prep). In addition to line cooling from H$_3^+$ and radiative cooling from CH$_4$, there are several other line cooling sources described in the following paragraphs. 
    
    We obtain the CO$_2$ line cooling by first calculating the escape probability, $P_{esc}$, for each cell, based on the values tabulated in \cite{kumer_co2_1974}. To then derive the density of excited CO$_2$ particles, n$_{CO_2-vib}$, we assume a steady state in accordance with Eq. 37 in \cite{johnstone_upper_2018}. The escape probability and density of excited CO$_2$ particles is subsequently put together with the Einstein coefficient, $A_{10}$, and the energy lost per emission at 15 $\mu$m to form
    \begin{equation}
        \Lambda_{CO_2} = 1.32510^{-13} n_{CO_2-vib} P_{esc}.
    \end{equation}
    
    Due to the vibrational transition at 5.3$\mu$m, line cooling from NO is calculated in a similar manner to CO$_2$ cooling described above. The density of NO$_{vib}$, n$_{NO_{vib}}$ is derived using the methods described in \cite{oberheide_impact_2013}, and is then used together with the Einstein coefficient, A$_{10}$, from the same source to form the expression for the cooling as
    \begin{equation}
        \Lambda_{NO_{vib}} = 3.75 A_{10}10^{-13} n_{NO_{vib}}.
    \end{equation}
  
    Line cooling from O is handled as described in  \cite{banks1973}, but only cooling due to emission at 63 and 147 $\mu$m is considered. This forms the expression for the cooling as
\begin{equation}
\Lambda_O =
\frac{
1.67 \cdot 10^{-18} e^{\left(-228/T_i\right)}
+ 4.59 \cdot 10^{-20} e^{\left(-326/T_i\right)}
}{
1 + 0.6 e^{\left(-228/T_i\right)}
+ 0.2 e^{\left(-326/T_i\right)}
}
n_O .
\end{equation}

    For cooling due to the emission from the Lyman $\alpha$ line, our calculations follow those of \cite{murray-clay_atmospheric_2009}, resulting in
    \begin{equation}
        \Lambda_{Ly-\alpha} = 7.5 10^{-19}  n_H  n_e  e^{ \frac{-118348.0}{T_i}}.
    \end{equation}
    Finally, H$_2$O cooling is expressed as \begin{equation}
        \Lambda_{H_2O} = n_H  n_{H_2O} L_{H_2O},
    \end{equation}
    where $n_H$, and $n_{H_2O}$ is the number density for H and H$_2$O respectively, and $L_{H_2O}$ is the line coefficient, calculated following the methods of \cite{KastingPollackH2O}.

\begin{figure}
    \centering
    \includegraphics[width=1.1\linewidth]{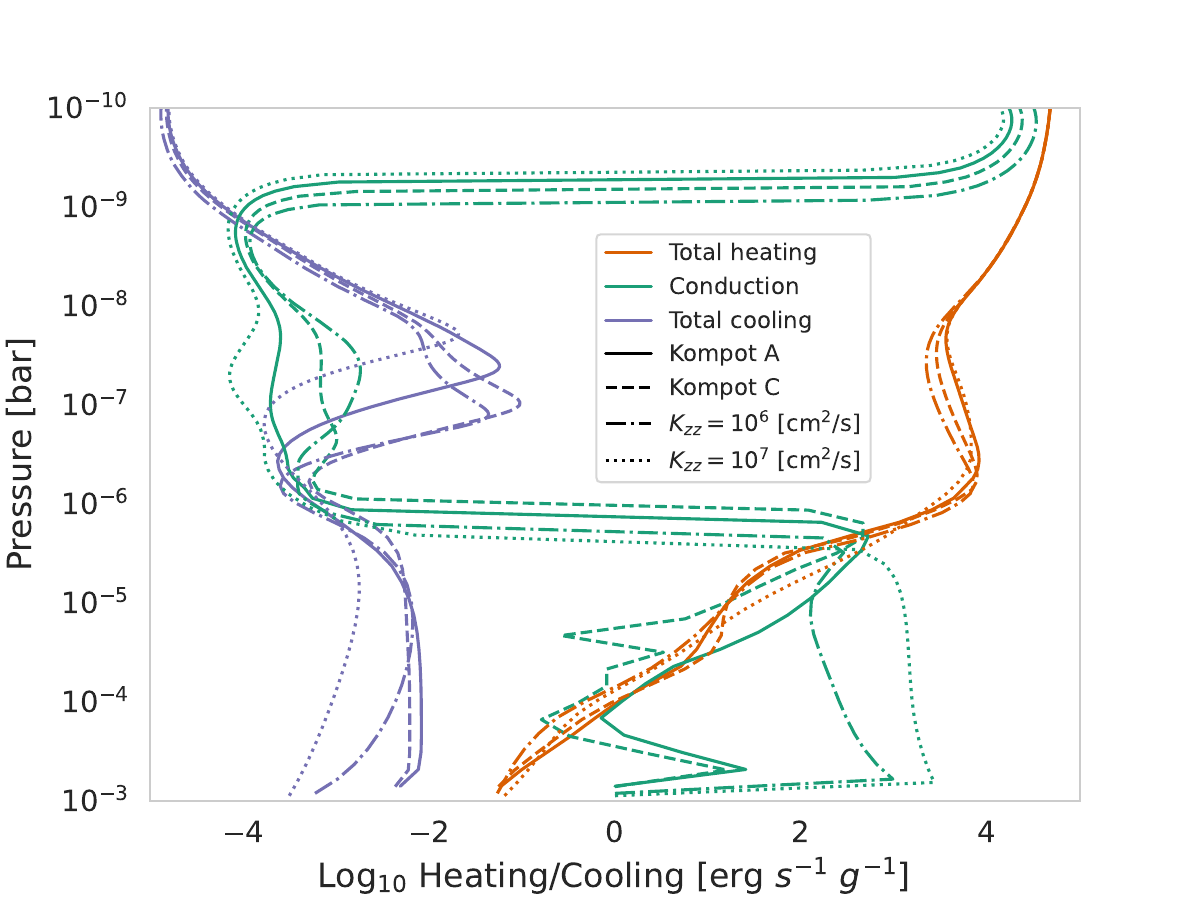}
    \caption{Total heating, cooling, and conduction for all four models. The cooling curves have been subjected to a sign change to illustrate the balance with conduction, where conduction on the left-hand side of 0 indicates cooling from conduction. Note that heat flux from the bottom boundary is not displayed in total heating, but shows up as a sharp feature in the conduction profile.}
    \label{fig:hccomp}
\end{figure}

\begin{figure}
    \centering
    \begin{tabular}{c}
        \includegraphics[width=0.48\textwidth]{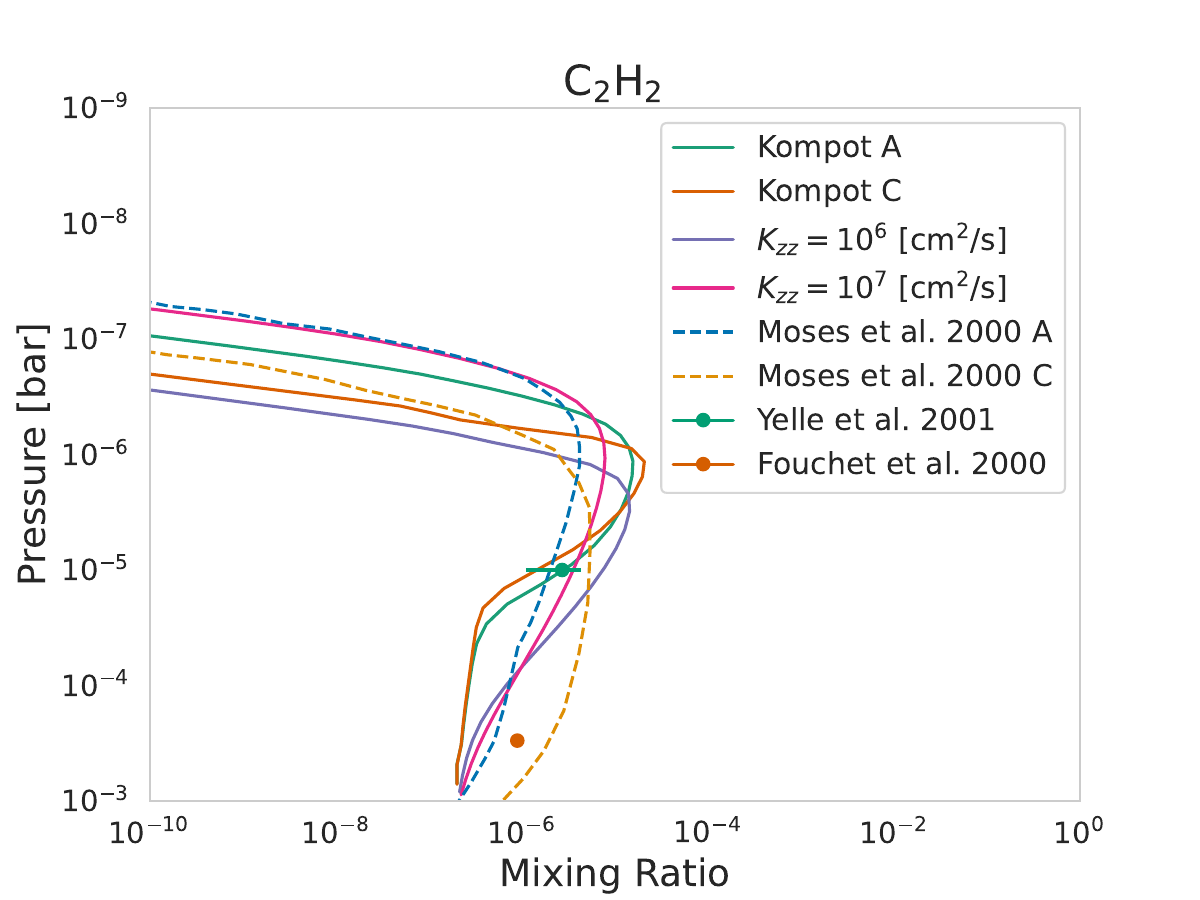} \\[2mm]
        \includegraphics[width=0.48\textwidth]{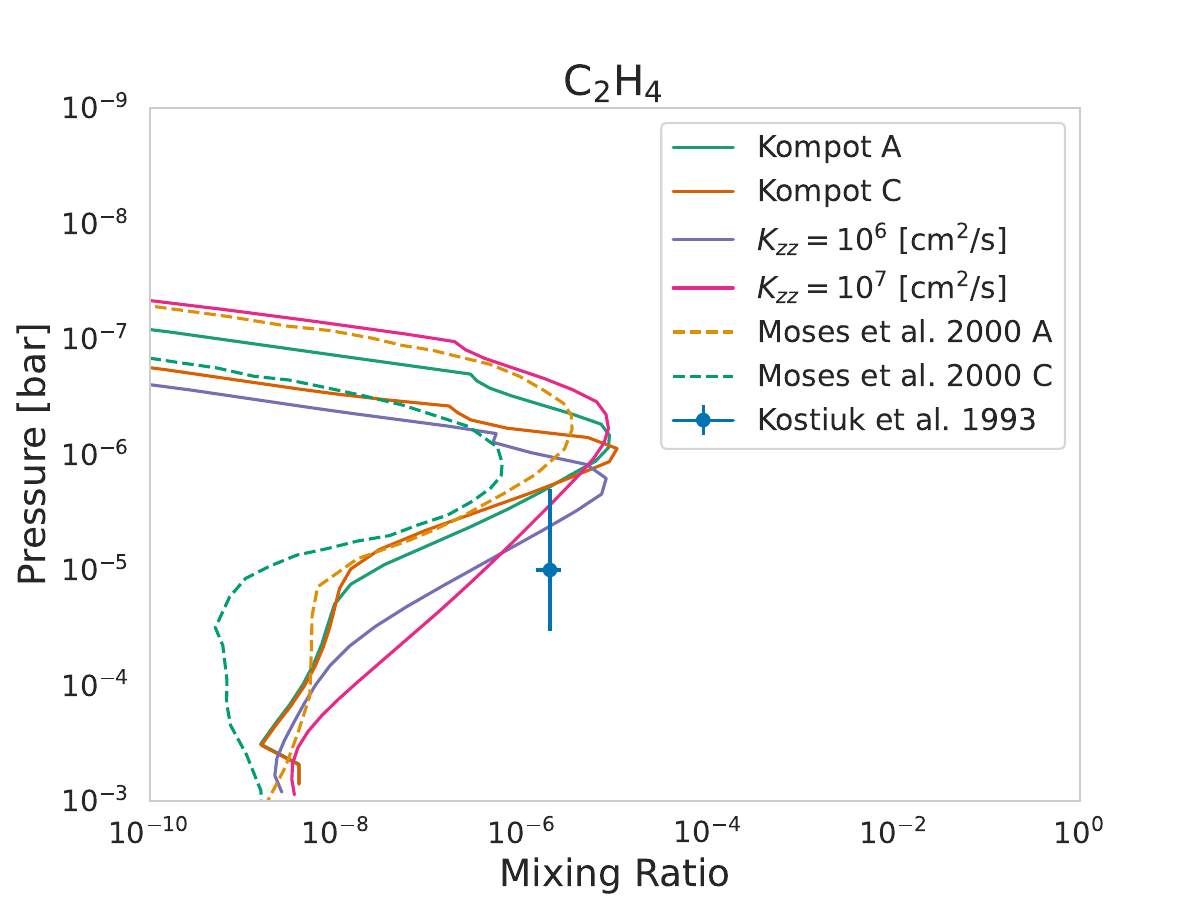}\\[2mm]
        \includegraphics[width=0.48\textwidth]{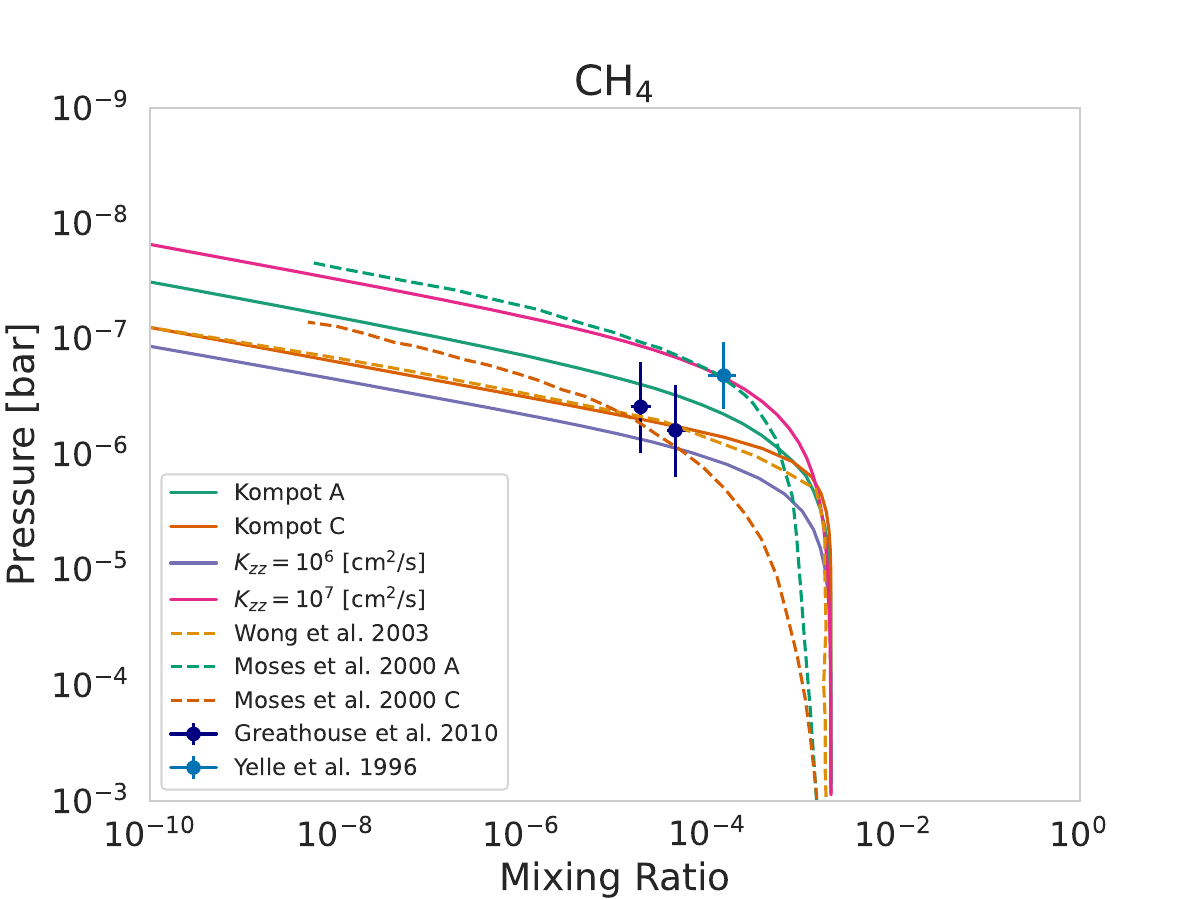}\\
    \end{tabular}
    \caption{Chemical profiles of C$_2$H$_2$, C$_2$H$_4$, and CH$_4$, together with measurements and the results from \cite{moses_photochemistry_2005}. The solid lines represent the results from the Kompot Code, and the dashed lines show results from other models. The error bars are as follows, \cite{yelle_structure_2001} are measurements by the IRSHELL (Infrared Cryogenic Echelle Spectrograph), \cite{fouchet_jupiters_2000} observed with ISO-SWS (Short Wave Spectrometer on the Infrared Space Observatory), \cite{kostiuk_temperature_1993} used IRHS (Infrared Heterodyne Spectrometer) at IRTF (NASA Infrared Telescope Facility), \cite{greathouse_new_2010} used the NH-UVS (New Horizons Alice UV Spectrometer), \cite{seiff_structure_1996} used in situ measurement from Galileo. Please note that the plots do not include the ghost cells at the bottom boundary, which are kept constant during the simulation.}
    \label{fig:minorchemmeas}
\end{figure}
\subsection{Diffusion}\label{sec:diffusion}
Diffusion in the model is governed by the balance of molecular and eddy diffusion, resulting in a diffusive flux: 
\begin{equation}
    \Phi_d = n v_d.
\end{equation}
This is calculated individually for each species, where $n$ is the number density of the species and $v_d$ is the diffusion speed given by
\begin{align}
v_d &= -D\Biggl[\frac{1}{n}\frac{dn}{dr} - \frac{1}{N}\frac{dN}{dr} 
      + \left(1-\frac{m}{M}\right)\frac{1}{p}\frac{dP}{dr} \nonumber
     +\frac{\alpha_T}{T}\frac{dT}{dr}\Biggr]\\
     & \quad - K_{zz}\left[\frac{1}{n}\frac{dn}{dr} - \frac{1}{N}\frac{dN}{dr}\right],
\end{align}
where $D$ is the molecular diffusion coefficient, $N$ is the total gas density, $r$ is the radial distance, $m$ and $M$ are the molecular mass of the considered species and the background gas, respectively, $P$ is the gas pressure, $\alpha_T$ is the thermal diffusion factor, and $K_{zz}$ is the eddy diffusion coefficient. We allow $K_{zz}$ to be a free input parameter in the code, which the user supplies as an altitude-dependent profile. 

The molecular diffusion coefficient, however, is not a free parameter. As in \citet{johnstone_hydrodynamic_2020} the value of $D$ for a specific species in a background gas is calculated as
\begin{equation}
    D = \frac{3}{2n(R + r_s)^2}\sqrt{\frac{k_BT}{2\pi}\left(\frac{1}{M} + \frac{1}{m}\right)},
\end{equation}
where $R$ is the background gas molecular radius, $r_s$ is the radius or the considered species, following Chapman-Enskog theory \citep{noauthor_mathematical_nodate}. 

Both the molecular and eddy diffusion coefficients are used not only to calculate the diffusive flux but also to determine the location of the homopause, which is important for initialising species densities. The homopause itself is found where the molecular diffusion coefficient is equal to that of the eddy diffusion coefficient, or $K_{zz} = D$. Below, in the homosphere, where the gas is well mixed, the scale height of the mixed gas is used to calculate the initial profile of species densities; the scale heights of individual species are then used in the heterosphere, where the gas is separated into distinct layers. In reality, the homopause is a transition region that does not exist at a single distance; however, since this sharp transition is only an initial condition, treating it as a point in our 1D model is a good approximation.

\subsection{Setup}\label{sec:input}

The above set of equations is solved on a finite grid of cells using the numerical solvers described in the appendices of \cite{johnstone_upper_2018}. The lower boundary is set to an altitude of 150km, and several ghost cells are added to the grid below the lower boundary for numerical reasons; these cells are not included in the figures presented in this paper. The upper boundary lies at the exobase, which is allowed to vary during the simulation in response to changes in the atmosphere's physical structure. The gas properties at the lower boundary that we assume are the neutral density, temperature, and mixing ratios of the major species. Additional parameters we set include the zenith angle, eddy diffusion coefficients, and the input solar XUV spectra, which are taken from \cite{CLARESPEC}. The input lower boundary conditions are described in Table \ref{tab:parameters} and Fig. \ref{fig:diff}. The mixing ratios at our input lower boundary in Table 1 are averages at 1 mbar from the model results of \cite{moses_photochemistry_2005}.

The simulation domain between the lower boundary and the exobase is initialised with a uniform temperature gas. The initial vertical profiles of the gas composition are determined by first finding the homopause where $K_{zz} = D$ and then assuming a uniform mixing ratio of the major species below this altitude and each species following its own mass-dependent scale height above this altitude. The Kompot model then evolves the gas properties within the simulation domain in response to the thermal and chemical processes considered, while keeping the lower boundary conditions constant, until the simulation reaches a steady state, which we take as our solution. All neutral, ion, and electron densities not presented in Table 1 are therefore created self-consistently by the photochemical model within Kompot.

Eddy diffusion is one of the major free parameters of the model, since it cannot be derived from physical principles or conditions alone \citep{lindzen_turbulence_1981, strobel_energy_1985}. We therefore consider four models that differ only in the assumed eddy diffusion coefficients. The eddy diffusion profiles, A and C, developed by Moses et al. (2005), served different purposes. Profile A was developed to fit the lower end of previously obtained number densities from both observations and models, and profile C was developed to fit the higher end. While profiles A and C feature slightly different chemistry in the original source, we use the same chemical network and mixing ratios for both models. Profile B is intentionally left out because it does not exhibit different chemistry in the original sources, and the coefficient values lie between profiles A and C. We additionally run models with two uniform values of the eddy diffusion coefficients. The vertical profiles of the eddy diffusion coefficients for each model are shown along with the molecular diffusion coefficients for CH$_4$ in Fig.\ref{fig:diff}.

\subsection{Transmission spectrum}

We calculate a transmission spectrum using 1D radiative transfer with the vertical temperature and chemical profiles from the Kompot simulation results of profile A.
We use the radiative transfer part of the \texttt{TauREx} code \citep{al-refaie_2021,al-refaie_2022}.
The radiative transfer forward model comprises 157 layers, spanning the pressure domain of the Kompot model ($p_\mathrm{BOA} \approx 7 \times 10^{-4}$ bar and $p_\mathrm{TOA} \approx 3 \times 10^{-12}$ bar).
We considered absorption cross-sections from the ExoMol project \citep{tennyson_2020,chubb_2021}, as well as the HITRAN \citep{gordon_2022} and HITEMP \citep{rothman_2010} archives.
The opacity sources contributing to the optical depth of the forward model include molecular absorption by \ce{CH4} \citep{yurchenko_2017}, \ce{CH3} \citep{adam_2019}, \ce{C2H2} \citep{chubb_2020}, \ce{C2H4} \citep{mant_2018}, \ce{CO} \citep{li_2015}, \ce{CO2} \citep{yurchenko_2020}, and \ce{H2O} \citep{polyansky_2018}, collision-induced absorption (CIA) from \ce{H2}--\ce{H2} \citep{abel_2011,fletcher_2018} and \ce{H2}--\ce{He} \citep{abel_2012} pairs, as well as Rayleigh scattering as included in \texttt{TauREx} \citep{cox_2015}.

\section{Results and discussion}

\subsection{Thermal structure}

In Fig. \ref{fig:mainPT}, we compare the thermal profiles of our four models with observational data and in situ measurements from Galileo \citep{yelle_structure_2001}. Our models are slightly colder than the measurements at higher pressures below the temperature inversion. This small difference could be due to either the absence of aerosol heating or potential non-LTE effects not included in the model. Around 10$^{-6}$ bar, just below the homopause, the thermal profile plateaus, followed by a thermal inversion. This structure arises because CH$_4$ is photolysed, thereby stifling its radiative cooling, heating the atmosphere. The upper atmosphere then heats rapidly until it is once again balanced by the emerging H$^3_+$ cooling, which becomes effective as ionisation increases and H$^3_+$ becomes available, radiating effectively in the infrared.

 Our model reproduces both the location of the thermal inversion and the temperature above it with good precision. Since Jupiter is a well-studied object with data from many different regions and sources, our model cannot agree with all measurements and models simultaneously. The spatial variation in temperature makes it difficult to represent accurately with a 1D model. Because Jupiter also has highly active auroral regions due to its strong magnetic field, it poses an additional challenge. As discussed earlier in Sec. \ref{sec:heating}, Kompot does not contain heating from particle precipitation or any other effects attributed to the magnetic field. In addition, upward-propagating waves and large-scale dynamics may transport energy into the upper atmosphere, thereby contributing to local temperature variability that Kompot currently cannot reproduce. It is also worth noting that the Sun's XUV spectrum is highly variable over time, adding another source of temporal variation.

 In Fig. \ref{fig:HCcomps}, we show the heating and cooling components for Model A. As seen on the left-hand side, Joule heating dominates throughout most of the atmosphere, accounting for $\sim 200$ TW when integrated over the total contribution, and is replaced around the homopause by chemical ($\sim 15$ TW) and XUV heating ($\sim 10$ TW). The physical processes behind this are all connected to the balance between gyrofrequencies and collision frequencies. Where they have similar values, the electrons can move perpendicular to the magnetic field, current flows, and we have resistive heating. Joule heating experiences the drop starting at around 1 $\mu$bar because the neutral density drops, causing the ion collision frequency $\nu_i$ to decrease, i.e. $\omega_{c_i} \gg \nu_i$. This reduces the denominator in Eq.\ref {eq:pedersen}, leading to a decrease in the Pedersen conductivity. 

In chemical and XUV heating, we see a drop of about an order of magnitude slightly below the homopause. In this region, the structure changes from well-mixed to layered, causing key absorbers, the heavy hydrocarbons, to stay at higher pressures. As a result, we observe that the heat released upon absorbing XUV radiation decreases. The sudden decline is then quickly recovered when atomic hydrogen, with its large photoabsorption cross-section, becomes more readily available.

The shape of the chemical heating profile has similar origins. At the homopause, key reactions for energy release turn off since ion-neutral chemistry is weakened because of the drop in number density and some reactants, such as CH$_4$, disappear faster with decreasing pressure than others. These reactions are quickly replaced by dissociative recombination, releasing energy as new hydrogen-based ions, such as H$^+$, H$_2^+$, and H$_3^+$, become available. Because total heating is independent of chemistry, since chemical heating is never the dominant heating source, it is largely unaffected by the eddy diffusion coefficients, as seen in Fig.\ref{fig:hccomp}.

Up until the upper parts of the homosphere (P $\sim 10^{-7}$ bar), cooling is heavily dominated by radiative cooling from hydrocarbons, present in this model, is cooling from CH$_4$, as detailed in section \ref{sec:cooling}. Since CH$_4$ is heavy, it has a small scale height, leading to the quick decrease in density just above the homopause. Together with photolysis that starts in the upper homosphere, these two effects are responsible for CH$_4$ no longer being available above the homopause. The deficit in CH$_4$ leads to less cooling because of the number density dependence, and as soon as $X$(H$_3^+)>X$(CH$_4$), line cooling from H$_3^+$ becomes the dominant cooling mechanism quickly in the right-hand side of Fig. \ref{fig:HCcomps}.

The balance between heating and cooling, together with conduction, can be seen in Fig. \ref{fig:hccomp}. Since conduction depends on the eddy and molecular diffusion coefficients, it will be affected. We observe this, for example, in the region from $10^{-3}$ to $10^{-5}$ bar, where the models differ most, leading to increased conduction in the two constant $K_{zz}$ models. The initial jump in conduction heating can explain why both of the constant $K_{zz}$ models are hotter in Fig. \ref{fig:mainPT}. This jump is a response to the lower boundary incoming heat flux, which accounts for around $\sim 200$ TW. In the same region, we also observe an increase in cooling from the constant $K_{zz}$ models, particularly with $K_{zz} = 10^7$ cm/s. The increased cooling is mainly a result of radiative cooling from CH$_4$, which exhibits a jump due to its temperature and number density dependence.

\begin{figure}
    \centering
    \begin{tabular}{c}
        \includegraphics[width=0.48\textwidth]{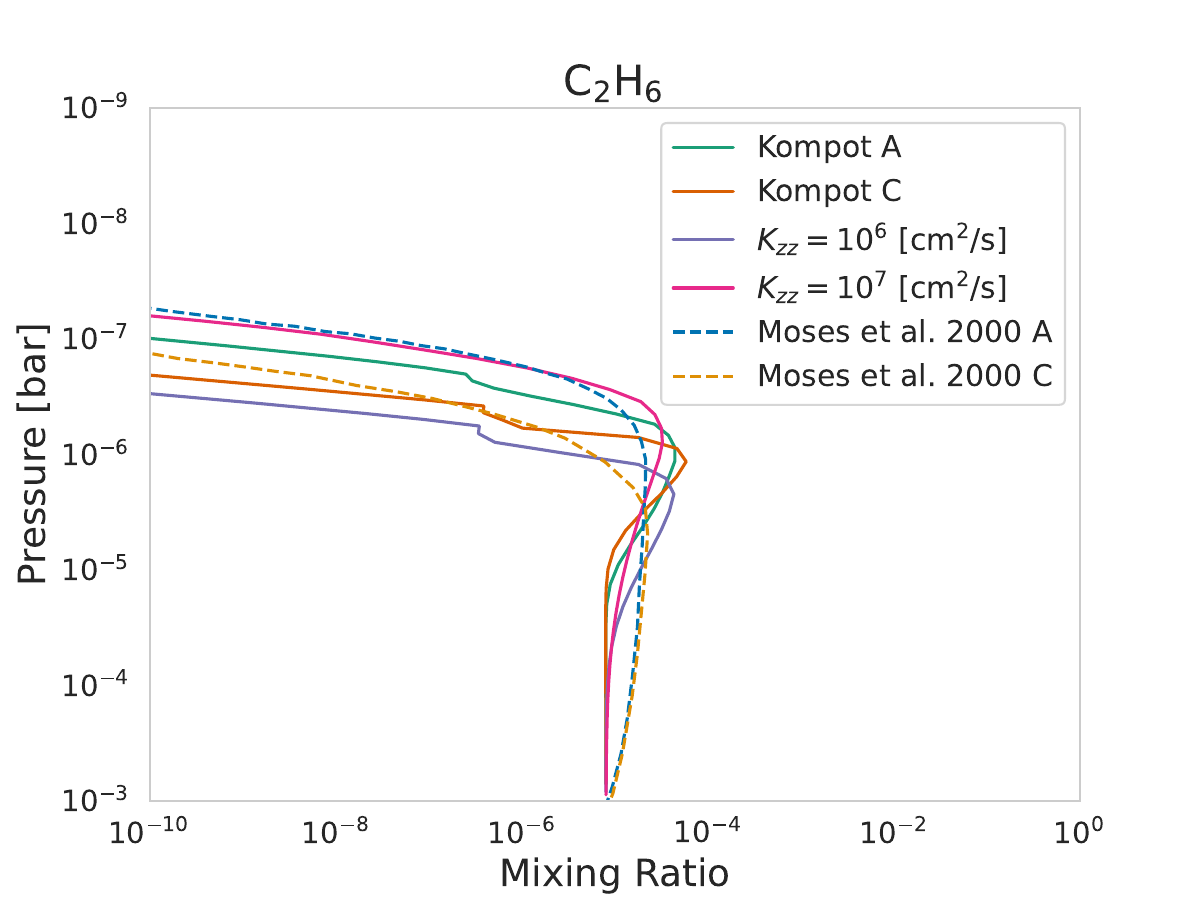} \\[2mm]
        \includegraphics[width=0.48\textwidth]{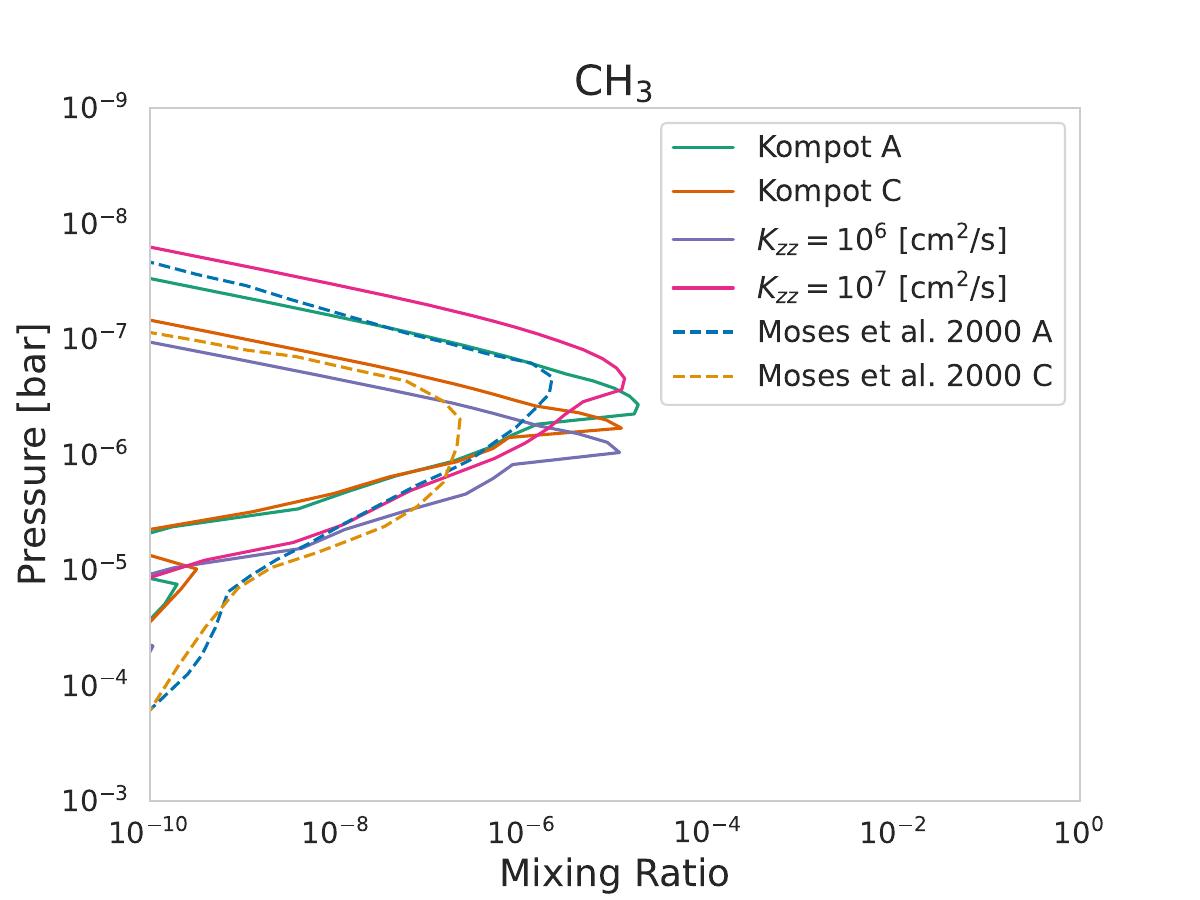} \\[2mm]
        \includegraphics[width=0.48\textwidth]{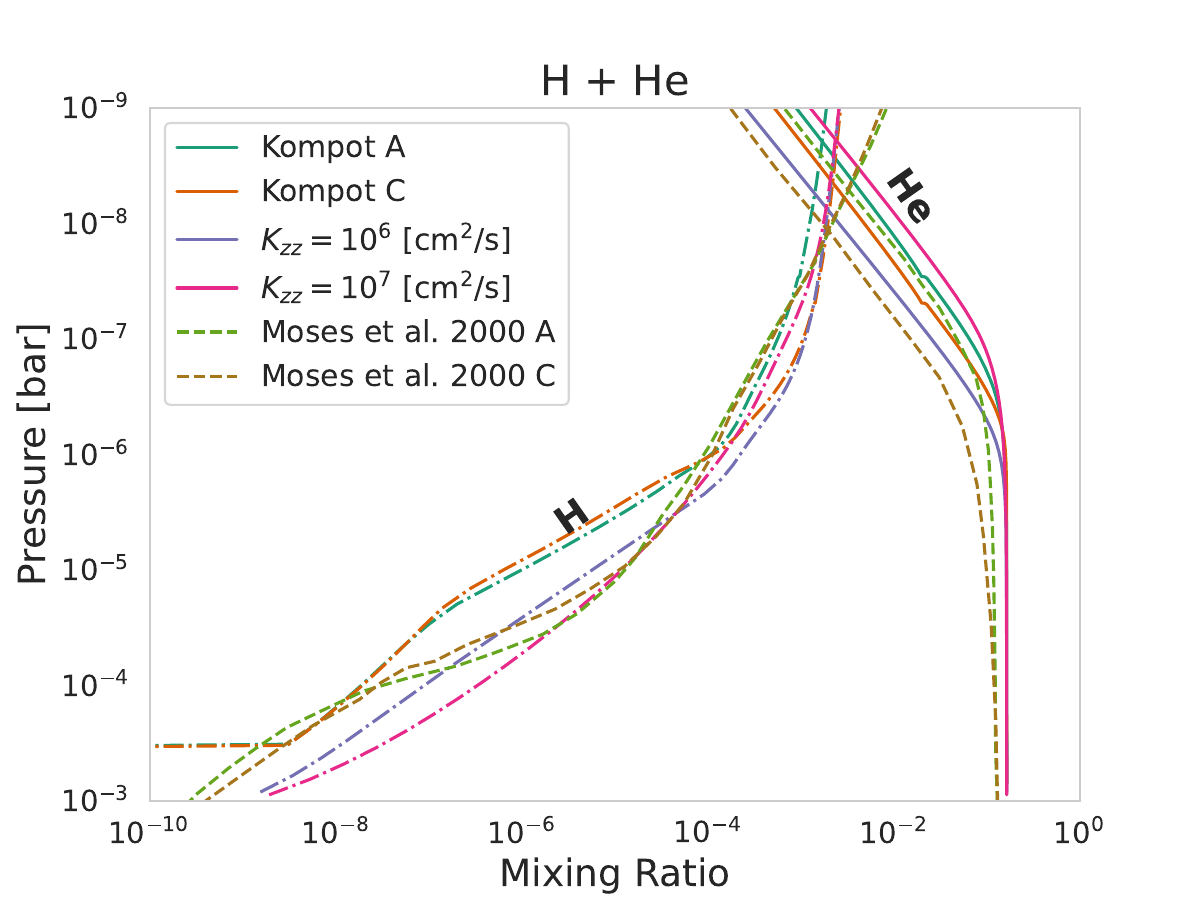} \\
    \end{tabular}
    \caption{Chemical profiles of hydrocarbons as well as H and He together with the model results from \cite{moses_photochemistry_2005}. Please note that the plots do not include the ghost cells at the bottom boundary, which are kept constant during the simulation.}
    \label{fig:minorchem}
\end{figure}

\subsection{Composition profiles}

Figures \ref{fig:minorchemmeas} and \ref{fig:minorchem} show the vertical mixing ratio profiles (without ghost cells) of the main hydrocarbons together with H and He, compared to the photochemical model of \cite{moses_photochemistry_2005} and available observations. There is good agreement with both other models and available observations, especially on crucial points such as the location of the homopause for CH$_4$ and the shapes of the profiles for minor hydrocarbons. 

To explain the shapes of the main hydrocarbon profiles shown, we first need to establish the role of CH$_4$ as the parent molecule. The main pathways to the C$_2$-bearing molecules are recombination reactions containing the radical CH$_3$. An abundance of CH$_3$ is initiated right below the homopause, due to the fast photolysis of CH$_4$ by solar UV photons. It then disappears because turbulent mixing is not strong enough to carry it higher, and because reactions with other hydrocarbons form additional hydrocarbons. This localised high mixing ratio of CH$_3$ helps the formation of heavier hydrocarbons around the homopause, and we see clear peaks in all of the C$_2$ bearing species in Fig. \ref{fig:minorchemmeas} and \ref{fig:minorchem}.

The atomic hydrogen abundance experiences a sharp rise up until the $\mu$bar level, indicating the onset of ion-neutral chemistry, eventually leading to the formation of H$_2^+$ and H$^+_3$. This is the effect of a combination of CH$_4$ photolysis and H$_2$ dissociation, which causes temperatures to drop at low pressures via infrared emission. As the helium abundance behaves as expected with altitude, i.e. staying constant until the homopause is reached, it can be used as a useful reference for how species unaffected by interactions with other species and only subject to diffusion behave.

\subsection{Influence of Io}
The main drivers of Jupiter's dynamics include planetary rotation, which is coupled with the magnetosphere via the conducting ionospheric layer in the upper atmosphere, the solar wind, and its moons, mainly Io \citep{al_saati_magnetosphereionospherethermosphere_2022}.

Together with the magnetic field lines from Jupiter and the plasma release from Io's volcanic activity, a plasma torus forms around Jupiter, which can be observed via multiple ionic emission lines in the extreme ultraviolet \citep{f_jupiter_nodate}. Moving on its orbit around Jupiter, Io also serves as a unipolar inductor \citep{Goldreich1969Io}, which accelerates some of the plasma from the torus towards the polar regions of Jupiter's ionosphere. These charged particles are accelerated to high energies and create footprints in the polar areas of Jupiter's ionosphere visible in X-rays and UV where they impact the planet (e.g. \citealp{Bhardwaj2000Aurora}). This particle precipitation changes the local conditions in the ionosphere due to energy deposition and additional ionisation, which has an effect on local electrical conductivity and Joule heating (e.g. \citealp{Bagenal_2020_IoAndEuropa}). 

Since the Kompot code is a 1D model, the extent of magnetospheric effects it accounts for is limited to the total Joule heating deposited in the atmosphere, which is parametrised. The code does self-consistently calculate the Pedersen conductivity, but does not include the external influences, such as precipitating particles. For this reason, additional analysis from magnetospheric models and observations would be required to better study Io's impact using our models.

\begin{figure}
    \centering
    \includegraphics[width=1.1\linewidth]{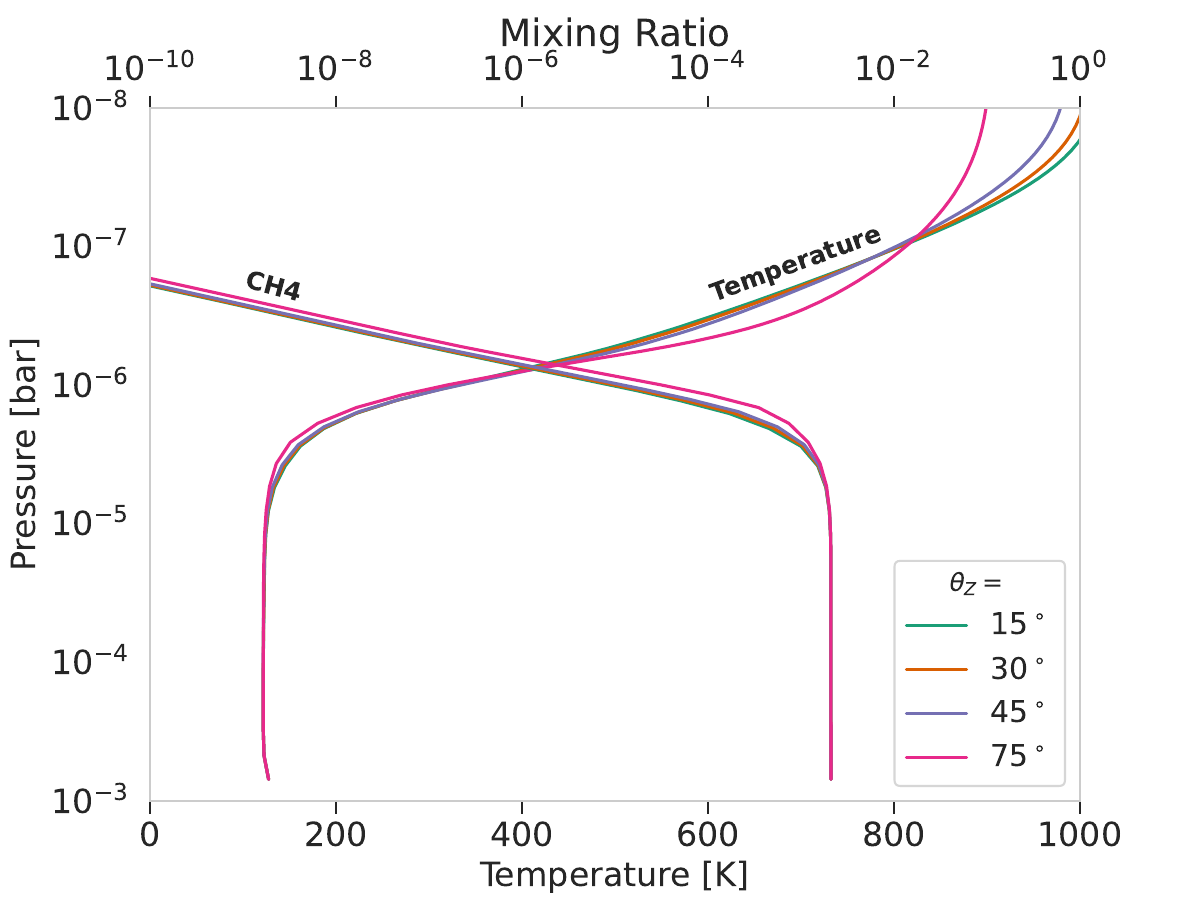}
    \caption{Comparison of the model sensitivity to different zenith angles. Impact on both mixing ratio profile of CH$_4$ and the thermal profile is shown for four different values of the zenith angle (coloured lines).}
    \label{fig:ZAcomp}
\end{figure}
\subsection{Sensitivity to free parameters}
Since lower boundary input molecular mixing ratios are the same in all models, constrained by several models and measurements \citep[][and references therein]{moses_photochemistry_2005}, we consider the free input parameters to be the eddy diffusion profile, total atmospheric Joule heating, and zenith angle. In Fig. \ref{fig:ZAcomp}, the results for different zenith angles with the other parameters fixed are shown. The only large-scale changes are in the top-most part of the atmosphere between $10^{-7}$ and $10^{-9}$ bars, where heating from XUV radiation dominates the energy balance. Globally, heating from XUV radiation is very minor compared to Joule heating, so the dominance of XUV heating comes from the absence of other effects. These differences are of the order of 100 K, still well within the limits of even the longitudinal deviations of Jupiter's temperature \citep{roberts_spatiotemporal_2025}. If we increase the total Joule heating by a factor of 2 beyond the values given in Table \ref{tab:parameters}, we observe an increase of approximately 100 K. We deem this acceptable given the spatial irregularity and uncertainty in the magnitude of the Joule heating present \citep{nishida_joule_1981, bougher_jupiter_2005, muller-wodarg_temperatures_2025}. 

\subsection{The role of eddy diffusion and the homopause}
The homopause plays a significant role in shaping chemical profiles. Where the gas is well mixed, mixing ratios of most of the initially available chemical species stay relatively uniform until the homopause is reached. Above the homopause, the species densities fall off according to their scale height, which means that the number density of heavier species decays faster. This is especially true for CH$_4$, for which photolysis reactions start just below its homopause. The combination of photolysis and the smaller scale height leads to the steep gradient we see in Fig. \ref{fig:minorchem}. 

A significant problem with eddy diffusion is that the coefficients vary substantially across objects and are notoriously difficult to constrain. In addition, they cannot be calculated analytically \citep{lindzen_turbulence_1981,strobel_energy_1985}. Both profiles used in this work are relatively similar in the lower pressure range considered up to about ~$10^{-5}$ bar, where they diverge and differ by at most an order of magnitude. 

Using constant eddy coefficients of $10^6$ and $10^7$ cm$^2$/s, we can illustrate the differences between a variable profile and a constant one and highlight the impact of an order-of-magnitude difference. The main divergence between the complex and constant K$_{zz}$ models visible in Fig. \ref{fig:diff}, is the consistent under- versus overestimation of the chemical abundances and temperature, compared to those of the more complex eddy coefficient profiles.  

\begin{figure}
    \centering
    \includegraphics[width=1.1\linewidth]{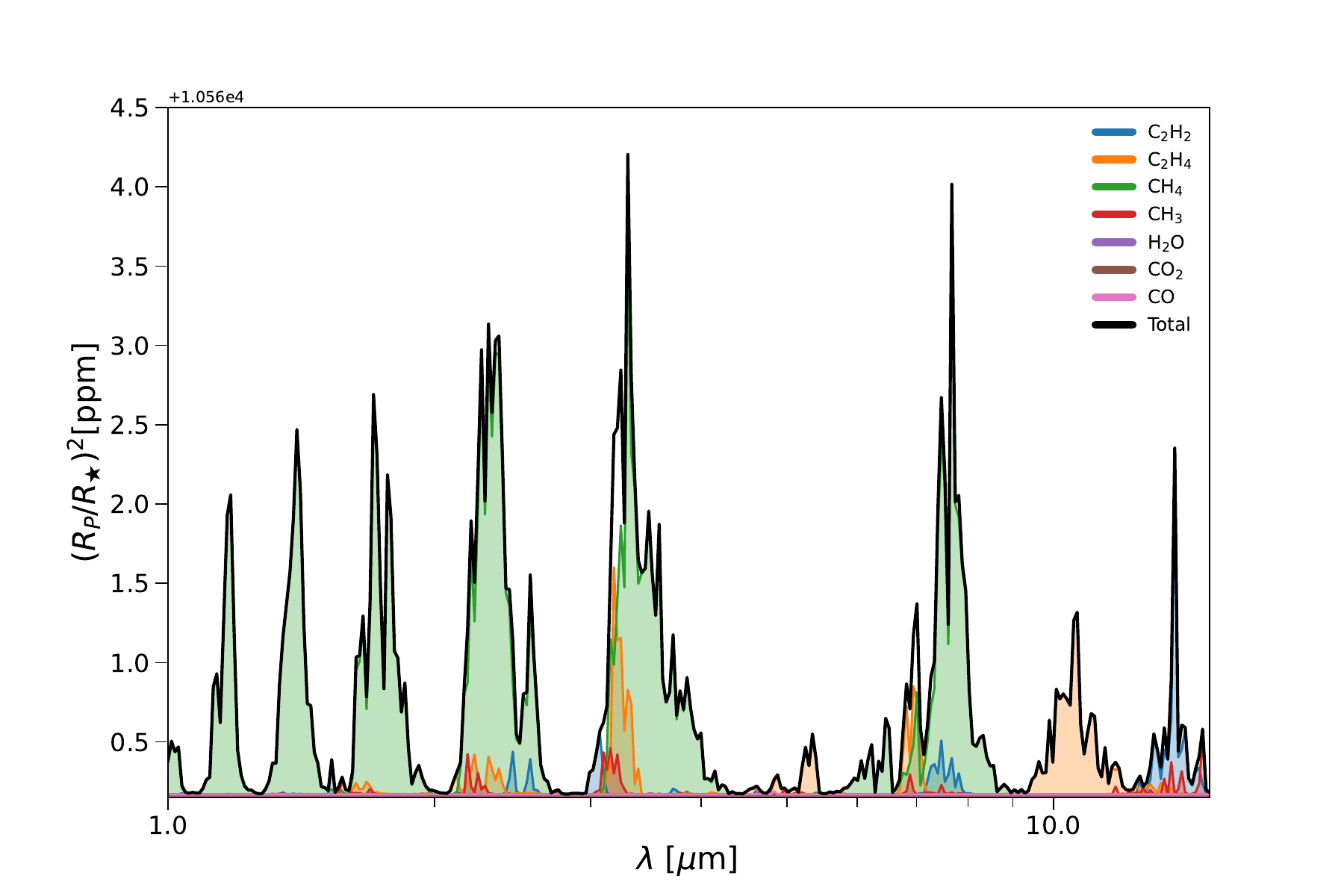}
    \caption{The transmission spectrum of Jupiter as seen between 1 and 15 $\mu$m calculated from Model A. Most of the contributions come from 0.1-1 mbar. The shaded regions describe absorption contributions from different species to the black line, which shows the total spectrum. Individual contributions from the CIA for H$_2$-H$_2$ and H$_2$-He, and from Rayleigh scattering, are intentionally omitted but still included in the total contributions. Please note that the y-axis values are shown relative to an offset of $+1.056\times 10^4$. }
    \label{fig:spec}
\end{figure}

\subsection{Spectral features}

Figure \ref{fig:spec} shows a synthetic transmission spectrum of Jupiter as an exoplanet, assuming it orbits a solar-analogue star, calculated from Model A. The spectrum spans wavelengths from 1 to 15 $\mu$m and is binned onto a logarithmically spaced wavelength grid with 300 points. The contribution to the spectrum comes predominantly from altitudes around $10^{-3}$ bar and up, as this is the lower boundary of our simulation, and we assume clouds are present below this level. Consequently, species primarily residing in the troposphere, specifically NH$_3$ and H$_2$O, are absent from this synthetic transmission spectrum.

In the wavelength range between 1 and 10 $\mu$m, CH$_4$ produces the most prominent spectral features \citep{1937ApJ....86..321W}, while C$_2$H$_2$ dominates at wavelengths longer than 10 $\mu$m. Beyond the dominant absorbers, the spectrum also contains several weaker features that observers could resolve in exoplanet observations of Jupiter, depending on the noise level and spectral resolution.

A minor C$_2$H$_4$ feature appears at 5.3 $\mu$m, but this feature would likely lie below the noise level. The C$_2$H$_4$ feature at 6.9 $\mu$m would be challenging to detect independently due to its proximity to a weaker CH$_4$ feature at a similar wavelength. However, a sufficiently high spectral resolution could resolve both features, in which case a pronounced peak would indicate the presence of both species. Another resolution-dependent feature appears at 3.2 $\mu m$, left of the strong CH$_4$ absorption at 3.7 $\mu$m, where sufficient spectral resolution could separate C$_2$H$_4$ from CH$_4$. While C$_2$H$_2$ may be difficult to detect between 1 and 10 $\mu$m, its strong feature at 13.7 $\mu$m could be detectable under favourable noise conditions.

To place these results in context, we compare our synthetic spectrum to available observations of Jupiter’s atmosphere. Near-infrared measurements obtained with XSHOOTER \citep{montanes-rodriguez_jupiter_2015} show strong CH$_4$ absorption between 1 and 2.5 $\mu$m, in good agreement with the dominant methane features in our model. More specifically, we replicate the relative strength of the absorption bands, with the 2.3 $\mu$m feature being the deepest and the subsequent 1.7 $\mu$m, 1.4 $\mu$m, and 1.2 $\mu$m showing diminishing depth. Similarly, \cite{kedziora-chudczer_modelling_2011} reports prominent CH$_4$ absorption bands between 1.6 and 1.9 $\mu$m, further supporting the robustness of methane as the primary absorber in this wavelength range. \cite{montanes-rodriguez_jupiter_2015} additionally report two absorption features at 1.5 and 2.0 $\mu$m, which they attribute to a layer of crystalline H$_2$O ice near the 0.5 mbar level. This pressure lies within our simulated domain, so its absence in our spectrum is not due to our lower boundary, as it is for NH$_3$ and tropospheric H$_2$O vapour. Rather, it reflects that our transmission spectrum calculation includes only gas-phase molecular opacity and does not account for aerosol or ice-cloud scattering. This suggests that a full accounting of Jupiter's near-infrared transmission spectrum, and by extension that of a cool gas giant exoplanet, requires aerosol microphysics in addition to gas-phase photochemistry and heating/cooling. Similarly, \cite{Irwin_Spec_2014} find that CH$_4$ dominates a synthetic Jupiter transmission spectrum over 0.4-15 $\mu$m, computed using radiative transfer modelling. Their spectrum also shows increasing depth with each band between 1-10 $\mu m$ in addition to a clear C$_2$H$_4$ feature around 7 $\mu m$. However, it is worth noting that the C$_2$H$_4$ feature at $\sim$ 10 $\mu m$ is missing in their absorption spectrum. 

A direct comparison with JWST data is difficult since only emission spectra are currently available \citep{rodriguez-ovalle_temperature_2024, melin_ionospheric_2024, harkett_thermal_2024}. We can, however, test our photochemistry directly by comparing the modelled abundances of C$_2$H$_2$ and C$_2$H$_6$ against retrievals spanning a range of latitudes and techniques: JWST/MIRI observations of the southern auroral region \citep{rodriguez-ovalle_temperature_2024}, disk-averaged Voyager/Cassini-CIRS retrievals \citep{nixon2007}, ground-based IRTF/TEXES global maps \citep{Fletcher2016}, and Gemini-N/TEXES polar profiles. Our modelled C$_2$H$_2$ profile mixing ratio peaks at $2\times 10^{-5}$ near $10^{-6}$ bar, which is a few times higher than in \cite{sinclair2023}, about twice as much as the normal retrieval from \cite{nixon2007}, but matching their fixed troposphere retrieval well. C$_2$H$_6$ shows best agreement with the normal retrieval from \cite{nixon2007}, and less peak agreement with their more turbulent fixed troposphere profile. The retrieved values of C$_2$H$_6$ from \cite{sinclair2023} also agree well in both shape and value with our model. This is consistent with the two species' differing photochemical lifetimes. C$_2$H$_6$ is comparatively inert, with a longer photochemical lifetime \citep{moses_photochemistry_2005}, making it well-mixed and largely insensitive to the local photochemical state. That our 1D model reproduces both species well across this spread of observations suggests that it is robust outside of regions with strong localised forcing.

\section{Summary and conclusions}

This work presents the first application of the Kompot model to Jupiter's upper atmosphere, extending a 1D first-principles framework to self-consistently simulate the average thermal and chemical structure between 10$^{-3}$ and 10$^{-10}$ bar. Treating Jupiter as an exoplanet orbiting a Sun-like star, we produce a globally averaged radial profile benchmarked against Galileo in situ measurements, JUNO/JIRAM retrievals, ground-based observations, and several models. We also present a synthetic transmission spectrum calculated between 1 and 15 $\mu$m using the TauREx radiative transfer code. With minor adjustments to the chemical network to fit the higher temperatures of extrasolar gas giants (planned for future work), the model can be extended to simulate the upper atmospheres of hot and warm Jupiters. The key takeaways are summarised below.

\begin{itemize}
    \item  Kompot performs well in reproducing average chemical and thermal profiles similar to measurements as well as photochemical models in the literature. Our model also reproduces the observed vertical distributions of the main hydrocarbons in the upper atmosphere and the location of the homopause across the models. 
    \item Eddy diffusion coefficients influence the overall profile of Jupiter's chemistry and thermal structure. The large differences highlight the importance of handling diffusion with care.
    \item According to our average model, the primary heating mechanism throughout most of the upper atmosphere of Jupiter is Joule heating. Close to the exosphere, at around $10^{-8}$ bar, it is overtaken by XUV driven heating. The main source of cooling below the homopause is radiative cooling by CH$_4$, which switches to H$_3^+$ line cooling in the heterosphere.
    \item  Assuming a thick cloud deck at 1 mbar, many of the spectral features observable in the IR and mid-IR other than CH$_4$ would be difficult to detect in transmission spectroscopy as CH$_4$ dominates the spectrum in this wavelength range. Some features of minor hydrocarbons would be detectable depending on resolution and noise levels. 
\end{itemize}

\begin{acknowledgements}
This work was carried out as part of the APPLE-Kompot ASAP project, supported by the Austrian  Forschungsf\"orderungsgesellschaft (FFG) funded through BMIMI.\end{acknowledgements}

\bibliography{references}
\bibliographystyle{aa} 

\end{document}